\documentstyle[12pt]{article}
\begin{document}

\begin{center}
{\large{\bf Screening in Anyon Gas}}\\[5mm]
Subir Ghosh\\[3mm]
{\sl Physics Department Dinabandhu Andrews College, Calcutta 700084, India}\\
\end{center}
\vspace{1cm}

\begin{abstract}
Anyon gas with interparticle (retarded) Coulomb interaction has been 
studied. The resulting system is shown to be a collection of dressed
anyons, with a screening factor introduced in their spin. Close
structural similarity with the Chern-Simons construction of anyons
has helped considerably in computing the screening effect. Finally
the present model is compared with the conventional Chern-Simons
construction.
\end{abstract}
\newpage

The possible existence of particles having arbitrary spin and statistics
has been proved quite some time back \cite{lm,w}. However, dynamical model
building, on the other hand, has proved to be rather controversial, with
the dispute still continuing. Basically there are two broad lines along
which the models are conceived: (i) The Chern-Simons (CS) construction
\cite{wb}, where a point charge is coupled to CS eletrodynamics. Removal
of the auxiliary (or statistical) CS gauge field renders the particle 
anyonic. (ii) The construction of minimal anyon field equations, 
where one starts
from very general physical postulates, such as the mass shell and 
Pauli-Lubanski condition for the particle \cite{j}. A variant of
the latter scheme is the spinning particle model \cite{cnp, sg, gor},
with which we are concerned in the present Letter. The connection between
the latter two is elaborated in \cite{gm}.

It is important to point out that individually both models represent
anyons. The controversy arises as regards the nature of the CS gauge
field in (i). The contention of \cite{wb}, that the only effect of
the CS gauge field is to influence the particle statistics and nothing
else, has been debated strongly in \cite{jsp}. Also the CS scheme fails
in the relativistic theory, relevant for cosmic string problems.

Our result in this Letter show conclusively that {\it anyons in the presence
of genuine interparticle
Coulomb interaction, are dressed as far as their fractional spin
(and hence the statistics parameter) is concerned}. The {\it screened} spin 
$S=\alpha j$ is
$\alpha=-{{Q^2}\over {16\pi^2\epsilon_0mc^2}}$, 
whereas in the spinning partcle
models \cite {cnp}, \cite {sg}, the spin is $S=j$, where $j$ is the Lagrangian spin parameter in
(\ref{eqlag}). Here $Q$ and $m$ are the charge and mass of the anyon, $c$
the velocity of light and $\epsilon_0$ a characteristic
property of the vacuum,
(to be elaborated later). The dimensionless quantity $\alpha$ is the
screening factor. We have considered a two particle system but
generalization to a many particle system is straightforward.

Also the other interesting feature of the model is its 
structural similarity
with CS construction \cite {w}. We show that in the slow moving and large
mass particle limit, the Coulomb field is structurally identical 
to the CS gauge field
solution in \cite {w}, with the identification of the
 CS $\theta$ parameter, $\theta=-{{4\pi\epsilon_0mc^2}\over j}$. 
The crucial difference lies in the qualitative nature of the CS gauge
field and the Coulomb field considered here. The former is sort of a
fictitous gauge field \cite {cwwh}, that couples to the fictitous charge
of the particle, whereas the latter is the real Coulomb field, responsible
for the Lorentz force between particles.
Apart
from this, there is the usual logarithmic Coulomb potential. Note that
in conventional CS scheme, the particles are endowed with the fractional
spin $S={{Q^2}\over{4\pi\theta_{CS}}}$ where $\theta_{CS}$ is the
arbitrary CS parameter and $Q$ is the fictional charge that couples
to the CS statistical gauge field. One recovers the logarithmic Coulomb
interaction as well \cite {cwwh}.
The above mentioned identification helps us to make use of the CS results
directly to compute explicitly the screening factor $\alpha$. The connection
between conventional CS scheme and our model will be elaborated at the end.
Although, some of the major results of this paper have appeared in \cite {gb},
the implications and consequences, discussed in the conclusion, were not 
emphasized before.

The spinning particle Lagrangian proposed by us in \cite {sg} is, ($c=1$),
\begin{equation}
L={\sqrt{m^2 u^2+{1\over 2}j^2\sigma^2+mj\epsilon^{\mu\nu\lambda}u_\mu
\sigma_{\nu\lambda}}}~,
\label{eqlag}
\end{equation}
where the velocity and canonical momenta are defined as
\begin{equation}
u^\mu={{dr^\mu}\over{d\tau}};~~~\sigma^{\mu\nu}=\Lambda_{\lambda}^{~\mu}
\dot{\Lambda}^{\lambda\nu}, 
\label{eqvel}
\end{equation}
\begin{equation}
P^\mu=-{{\partial L}\over{\partial u_\mu}};~~~S^{\mu\nu}=-{{\partial L}
\over{\partial \sigma_{\mu\nu}}}.
\label{eqmom}
\end{equation}
$(r^\mu,\Lambda^{\mu,\nu})$ is a Poincare group element, as well dynamical
variables with the property, $\Lambda\Lambda^T=\Lambda^T\Lambda=g$, where
$g$ is the Minkowski metric $g^{00}=-g^{11}=-g^{22}=1$.

The action in (\ref{eqlag}), $\int L d\tau$ is invariant under 
reparametrizations of the arbitrary parameter $\tau\rightarrow\tau'=f(\tau)$.
The details of the constraint analysis can be found in \cite{sg}. We will
use the relevant Dirac Brackets (DB) as and when necessary. Let us briefly
demonstrate the appearence of the arbitrary phase. The set of Second Class
Constraints (SCC)  and First Class Constraints (FCC) are
\begin{equation}
S^{\mu\nu}P_\nu\approx 0;~~~\Lambda^{0\mu}-{{P^\mu}\over m}\approx 0,
\label{eqscc}
\end{equation}
\begin{equation}
P^2 -m^2\approx 0;~~~\epsilon^{\mu\nu\lambda}S_{\mu\nu}P_\lambda-mj\approx 0.
\label{eqfcc}
\end{equation}
Let us transform the set of SCC's in (\ref{eqscc}) to strong equality.
The induced DB relevant to us,
$$
\{\Lambda^{\mu\nu}, S^{12}\}=(\Lambda^{\mu 1}g^{\nu 2}-
\Lambda^{\mu 2}g^{\nu 1})$$
\begin{equation}
+{1\over{m^2}}(P^\nu P^1\Lambda^{\mu 2}-P^\nu P^2\Lambda^{\mu 1}
-P_\rho\Lambda^{\mu\rho}P^1g^{\nu 2}+P_\rho\Lambda^{\mu\rho}P^2g^{\nu 1}),
\label{eqdb1}
\end{equation}
in the particle rest frame, $P^i=0,~P^0=m$, reduces to
$$
\{\Lambda^{11},S^{12}\}=\Lambda^{12};~~~\{\Lambda^{22},S^{12}\}=-\Lambda^{21};$$
\begin{equation}
\{\Lambda^{12},S^{12}\}=-\Lambda^{11};~~~\{\Lambda^{21},S^{12}\}=\Lambda^{22}.
\label{eqdb2}
\end{equation}
Using the rest frame $\Lambda$'s, ie. $\Lambda^{01}=\Lambda^{02}=0,~~
\Lambda^{00}=1$ we get the relations,
$$
\Lambda^{10}=\Lambda^{20}=0;~~(\Lambda^{12})^2+(\Lambda^{11})^2=(\Lambda^{21})^2
+(\Lambda^{22})^2=1,$$
\begin{equation}
\Lambda^{11}\Lambda^{21}+\Lambda^{12}\Lambda^{22}=0.
\label{eqrest}
\end{equation}
Hence in the reduced phase space we can parametrize the remaining
independent variables by,
\begin{equation}
\Lambda^{12}=cos\phi;~~\Lambda^{11}=sin\phi;~~S^{12}={{\partial}\over
{\partial\phi}},
\label{eqrot}
\end{equation}
where $S^{12}$ is the Pauli-Lubanski scalar in the rest frame,
$${{\epsilon^{\mu\nu\lambda}S_{\mu\nu}P_\lambda}\over m}\mid_{rest~frame}~=
{{S_{12}P_0}\over m}=S_{12}.$$
Also from counting the number of independent degrees of freedom in phase
space we see that out of three each (independent) $S^{\mu\nu}$ and
$\Lambda^{\mu\nu}$ variables one each of $S$ and $\Lambda$ remain, since
out of the set of six SCC's in (\ref{eqscc}) only {\it two} from each set
 (totalling four)
are independent. So far the FCC's have remained intact. This is exactly
the parametrization employed by Plyushchay in \cite {cnp}. This $\Lambda$
variable, (or equivalently $\phi$), gives rise to the arbitrary phase.
This is consistent with the fact \cite {pl} that specifically in 2+1-dimensions,
the number of (phase space) degrees of freedom for a particle with fixed mass
and spin is the same as that of a massive spinless particle. Here the
remaining degrees of freedom, ie. $\phi$ and $S^{12}$, can be removed by
choosing a gauge for the Pauli-Lubanski FCC. However, the effect of the 
spin variables present in (\ref{eqlag})
manifest itself in the non-trivial DB's, which gives rise to
the spin contribution in the total angular momentum.

Note an interesting departure in the constraint structure from the parent
3+1-dimensional model \cite {hr}, where the spin (FC) constraint appeared
as a combination of the SCC's $S^{\mu\nu}P_{\nu}\equiv 0$, due to some non-trivial
algebraic identities. The latter are absent in 2+1-dimensions, as a result
of which the spin constraint comes here independently, ie. it is not obtainable
from the other FCC and SCC's, in (\ref{eqscc}) and (\ref{eqfcc}).

Let us now move to the main body of our work. From now on we will use
rationalized MKS units and keep $c$ and $\hbar$ as they come. This will be
useful for the comparison of the final results and churning out numerical
estimates. We use the planar Coulomb law as
\begin{equation}
{\bf F}_{Coul}={{Q_1Q_2}\over{2\pi\epsilon_0 r}}{\bf n},
\label{eqcoul}
\end{equation}
where ${\bf r}=r{\bf n}$ is the separation between the charges $Q_1$ and $Q_2$,
and $\epsilon_0$ is the "permittivity" of the vacuum. ${\bf F_{Coul}}$ 
denotes the force between the particles.
This Coulomb law
is compatible with the Gauss law in a plane, 
$\nabla.{\bf E}={{\rho}\over{\epsilon_0}}$, where ${\bf E}$ and $\rho$ are
the electric field and charge density respectively.
 We introduce $\mu_0$ and $\epsilon_0$
as the "permeability" and "permittivity" of the vacuum, to keep the
relations same as their 3+1-dimensional counterpart. We only use the relation
$\epsilon_0\mu_0={1\over{c^2}}$. Denoting by $[O]$= dimension of $O$, we note
that 
$$[\epsilon_0]={{C^2}\over{M(L/T)^2}};~~
[\phi]={{M(L/T)^2}\over C};~~[A_i]={{M(L/T)}\over C}.$$
Here $M,~L,~T, ~C$ are mass, length, time and Coulomb. $\phi$ 
and $A_i$ are the scalar and vector potentials. We have the standard
relations,
$${\bf E}=-{\bf \nabla}\phi-\dot{\bf A},~{\bf B}=~{\bf \nabla}
X~{\bf A}.$$
Here ${\bf B}$ is the magnetic field.

We briefly show the construction of the relativistic Darwin Lagrangian for
a system of two interacting point charges in a plane. As the results have appeared
in \cite {gb}, we simply incorporate $c$, $\epsilon_0$ and $\mu_0$ in their
respective places. The retarded logarithmic potentials are,
\begin{equation}
\phi={1\over{2\pi\epsilon_0}}\int d^2r\rho({\bf r},t-{r\over c})ln{r\over {r_0}},
\label{eqscpot}
\end{equation}
\begin{equation}
{\bf A}={{\mu_0}\over{2\pi}}\int d^2r\rho({\bf r},t-{r\over c}){\bf v} ln{r\over{r_0}}.
\label{eqvpot}
\end{equation}
$r_0$ denotes some length scale where the potential due to a point charge
vanishes.
Expanding in terms of $v$-the particle velocity and keeping terms upto $O({{v^2}
\over{c^2}})$, with the charge density 
$\rho=Q~\delta({\bf r}-{\bf r}_{particle})$, we get
$$\phi={Q\over{2\pi\epsilon_0}}[ln{r\over{r_0}}-{1\over c}(rln{r\over{r_0}}\dot )
+{1\over{2c^2}}(r^2ln{r\over{r_0}}\ddot)],$$
$${\bf A}={{\mu_0}\over{2\pi}}Q{\bf v} ln{r\over{r_0}}.$$
Performing a gauge transformation,
$$\phi\rightarrow\phi'=\phi-{{\partial f}\over{\partial t}};~~~{\bf A}\rightarrow
{\bf A'}={\bf A}+\nabla f,$$
such that,
$${{\partial f}\over{\partial t}}={Q\over{2\pi\epsilon_0}}[-{1\over c}
(rln{r\over{r_0}}\dot)+{1\over{2c^2}}(r^2ln{r\over{r_0}}\ddot)],$$
$$\nabla f={Q\over{2\pi\epsilon_0}}(-{{\bf n}\over c}ln{r\over{r_0}}-{{\bf n}
\over c})+{Q\over{4\pi\epsilon_0c^2}}(2{\bf n}rln{r\over{r_0}}+{\bf r}\dot ),$$
the retarded potential $\phi$ is reduced to the standard Coulomb form,
\begin{equation}
\phi'={Q\over{2\pi\epsilon_0}}ln{r\over{r_0}},
\label{eqfsp}
\end{equation}
\begin{equation}
{\bf A}'=-{Q\over{2\pi\epsilon_0}}[{{\bf n}\over {c^2}}({\bf n.v}
+c(1+ln{r\over{r_0}}))+{{\bf v}\over {c^2}}].
\label{eqfvp}
\end{equation}
The interaction is simply of the minimal current-gauge field form
$J_{\mu}A'^{\mu}$, where $J_0=\rho=
Q\delta({\bf r}-{\bf r}_p)$, ${\bf J}=Q{\bf v}\delta({\bf r}-{\bf r}_p)$
and $A'^{\mu}$ is the above set, (\ref{eqfsp}) and (\ref{eqfvp}). Thus, to
$O({{v^2}\over{c^2}})$, the Lagrangian, or the Hamiltonian obtained just
below, incorporate the effect of Coulomb interaction between two charges,
taking into account the relativistic corrections via the retarded time.

The two-particle Darwin Hamiltonian is,
\begin{equation}
H={{p^2}\over m}-{{p^4}\over{4m^3c^2}}+{{Q^2}\over{2\pi\epsilon_0}}ln{r\over{r_0}}
+{{Q^2}\over{2\pi\epsilon_0c^2}}[{{\bf r.p}\over{mr}}({{\bf r.p}\over{mr}}
+c(1+ln{r\over{r_0}}))-{{p^2}\over{2m^2}}].
\label{eqdar}
\end{equation}
Note that the correction terms in ${\bf A'}$ are qualitatively different
from their 3+1-dimensional counterpart \cite{lan}. This has induced
difference in the Hamiltonian correction terms as well.

So far the effect of the particle spin has not been taken into account.
Now we do this via a non-canonical transformation. We rewrite 
(\ref{eqlag}),
\begin{equation}
L=c^2\sqrt{({{m^2u^2}\over{c^2}}+{{j^2\sigma^2}\over{2c^4}}+{{mj}\over{c^3}}
\epsilon^{\mu\nu\lambda}u_\mu\sigma_{\nu\lambda})}~,
\label{eqlc}
\end{equation}
with the dimensions of the phase space variables being,
$$[u]={L\over T};~~[\sigma]=[\Lambda\dot\Lambda]=T^{-1};~~[S_{\mu\nu}]
=[j]={{ML^2}\over T};~~[P_{\mu}]={{ML}\over T}.$$
With $P^2=m^2c^2$ and $S^2=2j^2$, the DB's relevant to us are \cite {sg},
\begin{equation}
\{r^{\mu},r^{\nu}\}=-{{S^{\mu\nu}}\over{m^2c^2}}=-{j\over{m^3c^3}}\epsilon
^{\mu\nu\lambda}P_{\lambda};~~\{r^{\mu},P^{\nu}\}=g^{\mu\nu};~~\{P^{\mu},
P^{\nu}\}=0.
\label{eqdb}
\end{equation}
Invoking the quantization prescription that 
${i\over {\hbar}}\{DB\}\rightarrow[commutator]$,
we arrive at the following commutators,
\begin{equation}
[r^{\mu},r^{\mu}]={{i\hbar}\over{m^2c^2}}S^{\mu\nu}={{i\hbar j}\over{m^3c^3}}
\epsilon^{\mu\nu\lambda}P_{\lambda};~~[r^{\mu},P^{\nu}]=-i\hbar g^{\mu\nu};
~~[P^{\mu},P^{\nu}]=0.
\label{eqqc}
\end{equation}

One can "solve" the algebra by introducing the non-canonical transformation
{\cite {cnp,gb}},
\begin{equation}
r^i=q^i+{j\over{m^2c^2}}\epsilon^{ij}p_j;~~P^i=p^i,
\label{eqncc}
\end{equation}
where $(q,p)$ constitute a canonical pair with the non-zero commutator
$[q^i,p^j]=-i\hbar g^{ij}$. The transformation simulates the spin
property of the particle, as it has originated from the non-trivial
$[r^i,r^j]$ commutator in (\ref{eqqc}), which was crucial in producing
the spin part of the total angular momentum. We will come to this point
again. Note that although we have a canonical position coordinate $q$, the
price to pay for this is that $q$ does not transform as a position vector.
However, this departure can be quite small for slowly moving heavy particles.

This modifies $H$ to,
$$H_{spin}=H(P^i=p_i,~~r^i=q^i+{j\over{m^2c^2}}\epsilon^{ij}p_j)$$
$$
={{p^2}\over m}-{{p^4}\over{4m^3c^2}}+{{Q^2}\over{2\pi\epsilon_o}}
[ln{q\over {r_0}}(1+\alpha)+\alpha(1+\alpha)+{j\over{mcq}}\beta(1-
2\alpha^2-\alpha ln{q\over{r_0}})$$
\begin{equation}
+({j\over{mcq}})^2\alpha^2\beta^2
-({j\over{mcq}})^2\alpha\beta^2],
\label{eqhs}
\end{equation}
where the two dimensionless variables $\alpha$ and $\beta$  are,
$$\alpha={{\bf q.p}\over{mcq}}\approx~O({v\over c});~~~\beta={{\epsilon
^{ij}q_ip_j}\over{mcq}}\approx~O({v\over c}).$$
Let us define,
$${\it A}^i={{Qj}\over{4\pi\epsilon_0mc^2q^2}}\epsilon^{ij}q_j
=\sigma{{\epsilon^{ij}q_j}\over{q^2}},$$
\begin{equation}
a^i=(1-\alpha ln{q\over{r_0}}-2\alpha^2){\it A}^i,
\label{eqcs}
\end{equation}
and rewrite $H_s$ as,
$$H_s={1\over m}({\bf p}-Q{\bf a})^2+{{Q^2}\over{2\pi\epsilon_0}}[ln{q\over
{r_0}}+\alpha(1+ln{q\over{r_0}})+\alpha^2-({j\over{mcq}})^2\alpha\beta^2 $$
\begin{equation}
+({j\over{mcq}})^2\alpha^2\beta^2]
-({{Qj}\over{4\pi\epsilon_0mc^2}})^2{{Q^2}\over{mq^2}}(1-\alpha ln{q\over{r_0}}
-2\alpha^2)^2.
\label{eqhsf}
\end{equation}

The identification \cite{gb} of our system with that of a point charge
interacting with Chern-Simons gauge field is now obvious. The $\alpha$-
independent term in $a^i$ is the explicit solution of the CS gauge field.
Hence we can identify \cite {j},
\begin{equation}
\theta=-{Q\over{\sigma}}=-{{4\pi\epsilon_0mc^2}\over j},
\label{eqtheta}
\end{equation}
where $\theta$ is the CS parameter in the CS 
Lagrangian,
$$L_{CS}={{c\theta}\over 2}\int d^2r\epsilon^{\mu\nu\lambda}\partial
_{\mu}{\it A}_{\nu}{\it A}_{\lambda}.$$
Also the magnetic flux connected to the charged particle is $\Phi$, where
\begin{equation}
\Phi=-{{Qj}\over{2\epsilon_0mc^2}}.
\label{eqflux}
\end{equation}
Note the $\Phi$ is of the proper dimension of magnetic flux. This
is one of 
our cherished results, where we have been able to obtain $\Phi$
in terms of the spinning particle parameters by simply borrowing the
CS result.

Let now elaborate on the previously advertised dressing induced by the
Coulomb interaction. Since we already identified our system with point charge
CS system, the results of the latter can be directly used. According to
CS theory {\cite {rj}}, the physical states can be shown to be carrying an
angular momentum eigenvalue $S$, which is related to the CS parameter
$\theta_{CS}$ by $S={{e^2}\over{4\pi\theta_{CS}}}$. Here $e$ is the
fictional charge of the particle that couples to the CS gauge field to
generate the anyon.
This is the well known
fractional spin. In our case,
\begin{equation}
S={{e^2}\over{4\pi\theta_{CS}}}=-{{Q^2j}\over{16\pi^2\epsilon_0mc^2}}.
\label{eqspin}
\end{equation}
Note that if $S=~s\hbar$, 
$$s=~{{e^2}\over {4\pi\hbar\theta_{CS}}}~
=~{e\over{2\theta_{CS}(h/e)}}=~{{\Phi}\over{4\pi\Phi_0}},$$
where $\Phi_0$ is the flux quantum and $\Phi$ is obtained from (\ref{eqflux}).

However, in case of the minimal spinning particle model {\cite
{cnp,sg}}, due to the non-trivial $[r^i,r^j]$ commutator, the angular
momentum is modified by the spin contribution in the following way,
\begin{equation}
J^{\mu}=-\epsilon^{\mu\nu\lambda}r_{\nu}p_{\lambda}-{j\over{\sqrt{p^2}}}p^{\mu}.
\label{eqam}
\end{equation}
Construction of the Pauli-Lubanski scalar ${\bf p.J}=-jmc$ shows clearly that
the particle spin is just $j$. Hence comparing with (\ref{eqspin}), we
notice the extra parameters or dressings that have appeared as a result of
the Coulomb interaction. This is the main result of the present work.

Let us now put the CS and our work in their proper perspectives. Our model
of interacting anyons can be cast in the following form,
$${\cal {L}}=[\Sigma_{i=1}^2({1\over 2}mv_i^2+e({\bf{v_i.\cal {A}}}-{\cal {A}}_0))
+{{c\theta_{CS}}\over 2}\int d^2r\epsilon^{\mu\nu\lambda}\cal {A}_{\mu}
\partial_\nu\cal {A}_{\lambda}]$$
\begin{equation}
+[\Sigma_{i=1}^2 Q({\bf {v_i.A}}-{\bf{A_0}})],
\label{eqlcs}
\end{equation}
where for simplicity we have used non-relativistic expressions. The
fictitous charge $e$ and gauge fiels $\cal {A}_{\mu}$ makes the
particles anyonic, and $Q$ and $A_{\mu}$ are their genuine charges and
Coulomb interaction. For consistency, when computing $A_{\mu}$, one
should consider the anyon spin as well. This will be complicated if
retardation effects are to be taken into account.

On the other hand, we have started with a spinless interacting system
with genuine charges and evaluated the (Darwin) Lagrangian with
retardation effects duly taken care of. Subsequently we turn the whole
system anyonic, (via (\ref{eqncc})), and obtain the screening effect.
The fact that our interacting anyon system, in the lowest order, is
structurally similar to the CS system, has made life easier, by allowing
us to borrow previous results.

We now conclude with the following comments:\\
(I) We have considered a system of interacting anyons, following our
spinning particle model and have shown that there is a screening effect
in the anyon spin, arising from mutual Coulomb interactions.\\
(II) We have shown how our system should be compared with the Chern-
Simons construction.\\
(III) We have not used the $c=~\hbar=~1$ convention and this has made
some of the relations look clumsy. We have persisted with this since all
the dimensions of electromagnetic quantities have been overhauled, as we
have taken the planar logarithmic Coulomb potential to be fundamental.\\
(IV) Unless there is a proper definition of planar $\epsilon_0$, with a
numerical value, it is of no use to speculate about numerical estimates.\\
(V) Finally, it would be interesting to see if the Chern-Simons
construction described above reproduces this screening.

\end{document}